\documentclass[aps,superscriptaddress,twocolumn,]{revtex4}
\usepackage{bm}
\usepackage{epsfig}
\usepackage{times}
\usepackage{bbm}
\usepackage{amssymb}
\usepackage{amsmath}
\usepackage{comment}
\usepackage{natbib}
\usepackage{color}
\usepackage{graphicx}
\usepackage[normalem]{ulem} 
\usepackage[loose,nice]{units}       
\usepackage{hyperref}                
\hypersetup{colorlinks=true,allcolors=blue}
\usepackage{hypcap}

\begin{document}
\title{The PESCADO Method for Autonomous Systems: An Application to Photoionization at Near-optical Wavelengths}

\author{S{\o}lve Selst{\o}}
\affiliation{Faculty of Technology, Art and Design, Oslo Metropolitan University, NO-0130 Oslo, Norway}

\author{Bendik Steinsv{\aa}g Dalen}
\affiliation{Faculty of Technology, Art and Design, Oslo Metropolitan University, NO-0130 Oslo, Norway}

\begin{abstract}
In a  recent publication, Dalen, and Selstø, Phys. Rev. A {\bf 111}, 033116 (2025), it was demonstrated how converged photo electron spectra
could be determined using a complex absorbing potential on a truncated numerical domain considerably smaller than the extension of the dynamical wave function.
That approach required simulation until virtually all unbound parts of the wave function was absorbed, far beyond the duration of the interaction with the external field. In this work we formulate the method in a semi-analytical manner which allows us to extrapolate to infinite times after the interaction with the external field.  
In addition to obtaining photoelectron spectra for hydrogen differential in energy and ejection angle, we also demonstrate how -- and when -- the absorber may be seen as a detector, distorting the angular distributions when the detector is placed in the extreme vicinity of the atom.
\end{abstract}

\maketitle

\section{Introduction}

In a series of papers, the method we have coined \textit{PhotoElectron Spectra calculated from Absorption via the Density Operator}, abbreviated \textit{PESCADO}, was developed \cite{Selsto2021, Selsto2022, Dalen2025}. It adds to an already rich plethora of methods where absorbers and other non-Hermitian interactions are applied in order to facilitate the study of photoionization on truncated numerical domains. One example of another such method is the \textit{mask method}, which combines a mask function with projections onto approximate scattering states in order to estimate photoelectron spectra~\cite{Chelkowski1998, Grobe1999, Lein2000, Tong2006, DeGiovannini2012, Tumakov2020, Mehmood2023}. Within a similar method, the one which has come to be known as tSURFF, such calculations are conveniently facilitated by recasting the working equations into surface integrals~\cite{Ermolaev1999, Ermolaev2000, Serov2001, Tao2012, Scrinzi2012}.
Both these methods rely on the projection onto Volkov states and, consequently, they fail to incorporate the Coulomb attraction from the atomic nucleus.

In general, obtaining the proper scattering states is often a non-trivial challenge. This is particularly so in situations when absorption takes place during interaction with the laser pulse and, consequently, the scattering states carry time-dependence. However, obtaining the proper, time-independent scattering states may also be non-trivial in situations in which the wave function is analyzed \textit{after} the interaction with a laser pulse. To this end, non-Hermitian interactions may often serve the additional purpose of facilitating the  calculation of photoelectron spectra from the wave function without detailed knowledge of the scattering states, see, e.g., Refs.~\cite{McCurdy2004, Palacios2007, Morales2016, Argenti2013}.

Ref.~\cite{Selsto2022} also adds to another bulk of literature which is more related to quantum foundations, namely the connection between non-Hermicity and detection~\cite{Kosloff1986, Kvaal2011, Tumulka2022}.
Studies addressing arrival times in quantum physics, in particular, often employ the notion of a complex absorbing potential (a CAP) acting as a detector, see, e.g., \cite{Allcock1969I, Allcock1969II, Allcock1969III, Muga2004, Ruschhaupt2004, Echanobe2008}. In most of these works, this correspondence is taken for granted without much justification nor any derivation from some underlying detector model. Ref.~\cite{Halliwell1999} is a noteworthy exception in this regard.
Also Ref.~\cite{Selsto2022} evades any direct correspondence between the CAP and any detector model. It does, however, provide insights on \textit{when} the correspondence between an absorber and a detector is an adequate one, and when it is not. By considering how information is extracted from absorbed waves during numerical simulation, it may be seen that the absorber, in effect, acts as a detector when the interaction is diagonal in the basis of projection, and that it does not act as a detector when it is non-diagonal.
Either situation comes with pros and cons -- depending on what is the purpose of the simulation.

In Ref.~\cite{Dalen2025} we presented converged photoelectron spectra for a hydrogen atom exposed to a laser pulse. We imposed a local CAP, i.e., one that is diagonal in position basis, and analyzed the absorbed part during and after interaction with the laser.
In the approach, converged photoelectron spectra were obtained using static scattering states, despite the fact that a significant part of the absorption happened before the laser pulse was over. This was done in the softly ultraviolet region. In the present work we study a similar case, albeit at longer photon wavelengths. And more importantly, we introduce an approach in which the dynamics is resolved in a semi-analytical manner after the interaction with the laser pulse. This approach, which is quite similar to what was done in Ref.~\cite{Morales2016}, which, in turn, is equivalent to approaches applied in \cite{Palacios2007}, evades the need to simulate the
system until virtually all of the liberated wave packed is absorbed. In addition to photoelectron energy distributions, we present and discuss angular distribution in two rather different contexts: In terms of the asymptotic ejection angle and in terms of the angle at which the electron is absorbed.
As the local CAP is diagonal in the projection basis in the latter case, we argue that it may be seen as detector placed in the extreme vicinity of the atomic nucleus.

In the next section, Sec.~\ref{Sec_Theory}, we will outline the method we implement. Firstly, we give a brief account of the PESCADO method for a dynamical system. While this is detailed elsewhere, it is worthwhile repeating here for convenience and context. Next, in Sec.~\ref{sec:PESCADO_Auto}, we explain how photoelectron spectra may be calculated by semi-analytical means when the Hamiltonian no longer carries time-dependence.
In Sec.~\ref{Sec_Explicit} we provide explicit formulas for the method's adaption to a calculation in which the wave function is expanded in Spherical Harmonics, while we,
in
Sec.~\ref{Sec_AngleResolved}, discuss angular distributions of the photoelectron specifically.
Finally, Sec.~\ref{Sec_Numerics} provides some details on the numerical implementation.
%
In Sec.~\ref{Sec_Results} we present physical results. These are mainly concentrated around a photon wavelength of $\lambda = 400$~nm, on the border between the optical and the ultraviolet regimes, while some results for $\lambda=200$~nm are also presented. Firstly, we focus on ionization probabilities doubly and singly differential in energy and asymptotic ejection angle, while, secondly, we go on to present angular distribution according to the angle in which photoelectrons are absorbed.
%
Conclusions are drawn in Sec.~\ref{sec:Conclusions}.
Atomic units are used where stated explicitly.

\section{Theory}
\label{Sec_Theory}

We solve the time-dependent Schrödinger equation,
\begin{equation}
\label{TDSE}
i \hbar \frac{\partial}{\partial t} | \Psi \rangle  = H_\mathrm{eff} | \Psi \rangle ,
\end{equation}
for a hydrogen atom exposed to a laser field of finite duration $T$ given by the vector potential ${\bf A}$. We will take this field to be homogeneous, i.e., we will apply the dipole approximation. The corresponding Hamiltonian reads
\begin{equation}
\label{HamHerm}
H = \frac{{\bf p}^2}{2m} - \frac{e^2}{4 \pi \epsilon_0} \frac{1}{r} + \frac{e}{m} {\bf p} \cdot {\bf A}(t) .
\end{equation}

In order to truncate the numerical domain, we augment this Hamiltonian with a local CAP,
\begin{equation}
\label{HwithCAP}
 H \rightarrow H_\mathrm{eff} = H - i \gamma.
\end{equation}
While in general not necessary, we will, as mentioned, take our CAP to be a local potential in this work. It will also be isotropic,
\begin{equation}
\label{LocalCAP}
\gamma = \gamma(r),
\end{equation}
where the CAP function $\gamma(r)$ is zero when the radial distance $r$ is smaller than the CAP onset $R_c$ and positive beyond.

\subsection{The PESCADO method for a non-autonomous system}
\label{PESCADO_nonAuto}

Waves which reach the region where $\gamma(r)$ is supported will gradually be attenuated. Thus, the introduction of the CAP will in effect impose absorbing boundary conditions -- ideally without introducing reflections. Correspondingly, whenever the wave function has a spatial overlap with the CAP, the norm of the wave function, or, equivalently, the trace of the density operator, will decrease. Specifically, between time $t$ and $t + \Delta t$, the pure-state density operator $\rho(t) = | \Psi\rangle \langle \Psi |$ will evolve into
\begin{equation}
\label{DepletionRho}
\rho(t+\Delta t) = \rho(t) -\frac{i}{\hbar}[H, \rho(t)] - \frac{1}{\hbar} \{\gamma, \rho(t) \} + O(\Delta t^2) .
\end{equation}
Correspondingly, the anti-commutator $\{\gamma, \rho\}/\hbar$ has been removed from the system's density operator.
While this part is no longer involved in the dynamics, nothing prevents us from \textit{analyzing} and \textit{aggregating} it. Suppose we are interested in the ionization probability differential in some continuous quantity $X$, this may be estimated by aggregating the corresponding diagonal elements of the part that has been removed from the density operator:
\begin{equation}
\label{DifferentialP_generic}
\frac{\partial P
}{\partial X} =
\frac{1}{\hbar} \int_0^t \langle X | \{ \gamma, \rho(t') \} | X \rangle \, \mathrm{d} t',
\end{equation}
where $|X\rangle$ are the corresponding eigenstates. This may be seen as a progressive projector valued measurement -- not for the actual density matrix $\rho$ but for the effective density operator given by the anti-commutator $\{\gamma, \rho(t) \}/\hbar$.

In the special case of momentum distributions, $|X \rangle = |\varphi_{\bf k} \rangle$, Eq.~(\ref{DifferentialP_generic}) would be the triply differential ionization probability
\begin{align}
\nonumber
&
\frac{\partial^3 P_\mathrm{ion}}{\partial^3 \textbf{k}} = \frac{1}{\hbar}  \int_0^t  \langle \varphi_\textbf{k} | \{ \gamma, \rho(t') \} | \varphi_\textbf{k} \rangle  \, \mathrm{d}t' =
\\ &
\label{dPdK}
\frac{2}{\hbar} \mathrm{Re} \; \int_0^t  \langle \varphi_\textbf{k} | \gamma | \Psi(t') \rangle \langle \Psi(t') | \varphi_\textbf{k} \rangle \, \mathrm{d}t' .
\end{align}
Here, $\hbar {\bf k}$ is the asymptotic momentum of the outgoing waves, and $\varphi_{\bf k}$ is the corresponding scattering state.

In Eq.~(\ref{dPdK}), the upper limit in the time integral, $t$, should be chosen large enough for virtually all outgoing waves to eventually be absorbed.

In situations in which the basis $\{ \varphi_\textbf{k} \}$ happens to be time-independent, the order of the time integration and the projections onto the relevant eigenbasis may be interchanged;
\begin{subequations}
\label{dPdXTimeless}
\begin{align}
&
\label{dPdXnoTime}
\frac{\partial^3 P_\mathrm{ion}}{\partial^3 \textbf{k}}
=
\frac{2}{\hbar} \mathrm{Re} \; \langle \varphi_\textbf{k} | \eta | \varphi_\textbf{k} \rangle
 \quad \text{where}
 \\ &
 \label{UpsDef}
\eta \equiv \gamma \int_0^t   | \Psi(t') \rangle \langle \Psi(t') | \, \mathrm{d}t' .
\end{align}
\end{subequations}
It would seem reasonable to assume that Eqs.~(\ref{dPdXTimeless}) are invalid if significant absorption happens during interaction with an explicitly time-dependent external field, in which case also the proper scattering states $\varphi_{\bf k}$ carry time-dependence. However, as was demonstrated in \cite{Dalen2025}, we may still achieve very accurate approximations using time-independent scattering states -- also when the system is subject to a time-dependent Hamiltonian.
While this may not appear very intuitive,
it is correct when the time-dependence of the dynamical scattering states resides exclusively in a global phase factor. As may be checked by inspection, such a phase factor cancels out of the expression of Eq.~(\ref{dPdK}). This, in turn, is due to the fact that a density matrix does not suffer from being defined only up to a global phase factor, as does the wave function. Correspondingly, applying time-independent scattering states in Eq.~(\ref{dPdK}) proves to be a reasonable approximation in many situations. And, in any case, a far better one than the much applied approach of approximating $\varphi_{\bf k}$ with time-dependent Volkov states, which neglect the Coulomb potential. We refer to Ref.~\cite{Dalen2025} for more detail in this regard.

As mentioned, the upper time-integral in Eqs.~(\ref{DifferentialP_generic}), (\ref{dPdK}) and (\ref{dPdXTimeless}) should set be large enough to allow for absorption of virtually all outgoing waves. With near-zero energy components, this would require extrapolation towards infinity. In a numerical simulation we must, however, settle for a finite time. Choosing this finite upper time limit would require a trade-off between computational effort and accuracy in the photoelectron spectra -- specifically at low energies. Alternatively, in the case of an atom exposed to a laser pulse of finite duration $T$, we may exploit the fact that the system is time-independent (autonomous) for $t>T$. Hopefully, this notion should also explain the seemingly self-contradictory title of the present work.

\subsection{The PESCADO method for an autonomous system}
\label{sec:PESCADO_Auto}

Suppose we have aggregated outgoing waves throughout the duration of the laser pulse to obtain
\begin{equation}
\label{AggregateUntilT}
\eta_T \equiv \gamma \int_0^T   | \Psi(t) \rangle \langle \Psi(t) | \, \mathrm{d}t .
\end{equation}
When inserted into Eq.~(\ref{dPdXnoTime}), this provides a partial contribution to the photoelectron spectrum. The remaining contribution, which we will coin $\eta_\infty$, can be determined without integrating the time-independent
Schr{\"o}dinger equation numerically up to some large but finite final time.
Instead, we may describe the time-dependent wave function,
\begin{equation}
\label{StateBeyond}
|\Psi(t>T) \rangle = \exp(-i H_\mathrm{eff}^{(0)} (t-T)/\hbar) \, | \Psi(t=T) \rangle ,
\end{equation}
by semi-analytical means.
Here $H_\mathrm{eff}^{(0)}$ is the time-independent part of the effective, non-Hermitian Hamiltonian, i.e., Eq.~(\ref{HwithCAP}) with ${\bf A}={\bf 0}$ in Eq.~(\ref{HamHerm}).
We diagonalize our effective Hamiltonian, Eq.~(\ref{HwithCAP}), expand our state at time $t=T$ in these eigenstates and multiply each component with the proper time-factor:
\begin{equation}
\label{StateBeyondExpand}
|\Psi(t>T) \rangle = \sum_{n} e^{-i \epsilon_n (t-T)/\hbar} c_n | \phi_n \rangle ,
\end{equation}
%
where $| \phi_n \rangle$ and $\epsilon_n$  are the right eigenstates and eigenenergies, respectively, of $H_\mathrm{eff}^{(0)}$:
\begin{equation}
\label{Eigenstate_Heff}
H_\mathrm{eff}^{(0)} | \phi_n \rangle = \epsilon_n | \phi_n \rangle .
\end{equation}
We assume this spectrum to be well approximated by a countable set of pseudo-continuum states as opposed to a truly continuous spectrum. The coefficients $c_n$ in Eq.~(\ref{StateBeyondExpand}) are the expansion coefficients for the the wave function at time $t=T$,
\begin{equation}
\label{ExpansionNonOrth}
|\Psi(t = T) \rangle = \sum_n c_n | \phi_n \rangle .
\end{equation}
Since $H_\mathrm{eff}^{(0)}$ is not Hermitian, the eigenenergies $\epsilon_n$ may be complex, $\{ |\phi_n \rangle \}$ does not constitute any orthonormal basis, and the expansion coefficients, $c_n$, are \textit{not} the usual projections, $c_n \neq \langle \phi_n | \Psi(T) \rangle$.
However, together with the left eigenstates or, correspondingly, the right eigenstates of $\left( H_\mathrm{eff}^{(0)} \right)^\dagger$, which we will label with a tilde,
\begin{equation}
\label{LeftEigenStates}
\langle \tilde{\phi}_n | H_\mathrm{eff}^{(0)} = \epsilon_n \langle \tilde{\phi}_n | ,
\end{equation}
a bi-orthonormal set can be constructed:
\begin{equation}
\label{BiOrthogonal}
\langle \tilde{\phi}_n | \phi_m \rangle =
\langle  \phi_n | \tilde{\phi}_m \rangle = \delta_{n,m} .
\end{equation}
Note that mathematically, the complex eigenenergies $\epsilon_n$ coincide for the left and right eigenstates. By virtue of Eq.~(\ref{BiOrthogonal}), the states $|\tilde{\phi}_n \rangle$ may be calculated numerically by inverting the matrix consisting of the $|\phi_n \rangle$ states; obtaining any additional set of eigenstates by actually diagonalizing $\left( H_\mathrm{eff}^{(0)} \right)^\dagger$ is not required.

With Eq.~(\ref{BiOrthogonal}) the expansion coefficients in Eq.~(\ref{ExpansionNonOrth}) are found as
\begin{equation}
\label{CalculateCoeffs}
c_n = \langle \tilde{\phi}_n | \Psi(t=T) \rangle .
\end{equation}

Now, with the spectral representation, Eq.~(\ref{StateBeyondExpand}), we may determine the contribution to $\eta$ in Eq.~(\ref{UpsDef}) from time $t= T$ to $t \rightarrow \infty$:
\begin{align}
\nonumber
\eta_\infty  & = \gamma \int_T^\infty
\sum_n c_n
e^{-i\epsilon_n (t-T)/\hbar} | \phi_n\rangle
\\
\label{AnalyticalEta}
\times &
\sum_{n'} \langle \phi_{n'} |
c_{n'}^* e^{+i \epsilon_{n'}^*(t-T)/\hbar} \, \mathrm{d}t.
\end{align}
The time-integral is readily determined:
\begin{equation}
\label{EtaIntegrated}
\eta_\infty = -i \sum_n \sum_{n'} \frac{c_n c_{n'}^*}{\epsilon_n - \epsilon_{n'}^*} \gamma | \phi_n \rangle \langle \phi_{n'} |  .
\end{equation}
Here we have used the fact that the imaginary component of $\epsilon_n$ should be non-positive for all $n$. This holds because the CAP constitutes a positive semidefinite operator, $\gamma \geq 0$~\cite{Bendixson1902, Wielandt1955}.
It is also crucial that the eigenenergies are only real to the extent that the corresponding eigenstates do not overlap with the CAP:
\begin{equation}
\label{NoOverlap}
\epsilon_n \in \mathbb{R} \Rightarrow \gamma | \phi_n \rangle = 0 .
\end{equation}
Consequently, only terms for which $\epsilon_n$ has a strictly negative imaginary component contribute to the sum in Eq.~(\ref{EtaIntegrated}). This, in turn, ensures that the integral in Eq.~(\ref{AnalyticalEta}) actually converges as the denominator in the terms in Eq.~(\ref{EtaIntegrated}), $\epsilon_n-\epsilon_{n'}^*$, never vanish. Thus, from a mathematical point of view, we are allowed to restrict the terms to include only those for which $\mathrm{Im} \, \epsilon_n <0$:
\begin{equation}
\label{EtaIntegrated_v2}
\eta_\infty = -i \gamma \sum_{\mathrm{Im} \epsilon_n <0} \sum_{n'} \frac{c_n c_{n'}^*}{\epsilon_n - \epsilon_{n'}^*} | \phi_n \rangle \langle \phi_{n'} | .
\end{equation}
While mathematically justified, see, e.g., Ref.~\cite{Morales2016}, it is not obvious whether this notion should be taken literally \textit{numerically}. We will return to this issue.
For the sake of clarity we emphasize that the restriction $\mathrm{Im} \, \epsilon_n < 0$ only applies to the sum over $n$, not the sum over $n'$.

Finally, the differential ionization probability is the coherent sum of contributions aggregated during interaction with the laser pulse and contributions
picked up afterwards:
\begin{align}
\nonumber
&
\frac{\partial^3 P_\mathrm{ion}}{\partial^3 {\bf k}}
=
\frac{2}{\hbar} \, \mathrm{Re} \, \langle \varphi_{\bf k} | \left( \eta_T + \eta_\infty \right) | \varphi_{\bf k} \rangle = \\
&
\label{SumBeforeAfter}
\left. \frac{\partial^3 P_\mathrm{ion}}{\partial^3 {\bf k}} \right|_\mathrm{before} +
\left. \frac{\partial^3 P_\mathrm{ion}}{\partial^3 {\bf k}} \right|_\mathrm{after} ,
\end{align}
where
\begin{subequations}
\label{DefBeforeAfter}
\begin{align}
\label{DefBefore}
& \left. \frac{\partial^3 P_\mathrm{ion}}{\partial^3 {\bf k}} \right|_\mathrm{before} =
\frac{2}{\hbar} \mathrm{Re} \, \langle \varphi_{\bf k} | \eta_T | \varphi_{\bf k} \rangle  \quad \text{and}
\\
\label{DefAfter}
& \left. \frac{\partial^3 P_\mathrm{ion}}{\partial^3 {\bf k}}
\right|_\mathrm{after} =
\frac{2}{\hbar} \mathrm{Re} \, \langle \varphi_{\bf k} | \eta_\infty | \varphi_{\bf k} \rangle =
\\ \nonumber &
\frac{2}{\hbar} \, \mathrm{Im} \,
 \sum_{\mathrm{Im} \epsilon_n <0} \sum_{n'} \frac{c_n c_{n'}^*}{\epsilon_n - \epsilon_{n'}^*} \langle \varphi_{\bf k} | \gamma | \phi_n \rangle \langle \phi_{n'} | \varphi_{\bf k} \rangle .
\end{align}
\end{subequations}
We emphasize that the sum in Eq.~(\ref{SumBeforeAfter}), despite a choice of notation which may appear somewhat misleading, indeed \textit{is} a coherent sum; the two terms come with signs. This would explain why conventional projections, $\left| \langle \varphi_{\bf k}|\Psi(T)\rangle \right|^2$, cannot replace the right hand side of Eq.~(\ref{DefAfter}).

\subsection{Explicit expressions in a spherical expansion}
\label{Sec_Explicit}

It is quite common in numerical simulations of photoionization to express the angular dependence of our wave function via an expansion in Spherical Harmonics:
\begin{equation}
\label{ExpandSpher}
\Psi(r, \Omega; t) \approx \frac{1}{r} \sum_{\ell=0}^L \sum_{m=-\ell}^\ell f_{\ell, m}(r; t) Y_{\ell, m}(\Omega) .
\end{equation}
The implementation used in the numerical examples presented here is no exception. We will write out explicit working equations for wave functions given this way.

For a system with cylindrical symmetry, such as an atom with a spherically symmetric potential interacting with a linearly polarized laser field within the dipole approximation, the magnetic quantum number $m$ does not change. In the following, we will take it to be fixed and, for the most part, omit this indexing.
With this symmetry, the indexing of the eigenstates -- and their eigenenergies -- are shifted according to
\begin{equation}
\label{ShiftIndex}
\phi_n \rightarrow \phi^{\ell}_{n_\ell} \quad \text{and} \quad
\epsilon_n \rightarrow \epsilon^{\ell}_{n_\ell} .
\end{equation}
As indexing an index becomes rather cumbersome notationwise, we will also omit the $\ell$ index on $n_\ell$ in the following and take it to be implicit.


The differential ionization probability expressed in terms of (real) energy $\varepsilon$ and asymptotic ejection angle $\Omega_k$,  instead of the wave vector ${\bf k}$, reads
\begin{equation}
\label{DoublyDiffEnergyAngle}
\frac{\partial^2 P_\mathrm{ion}}{\partial \varepsilon \partial \Omega_k} = \frac{2}{\hbar} \mathrm{Re} \, \langle \varphi_{\varepsilon, \Omega_k} | \eta | \varphi_{\varepsilon, \Omega_k} \rangle .
\end{equation}
In the hydrogen case, the scattering states are
\begin{subequations}
\label{CoulombScattering}
\begin{align}
\label{ScatteringCoulomb}
& \varphi_{\varepsilon,\Omega_k}(r, \Omega) = \frac{1}{r} \sum_{\ell} i^\ell e^{i \sigma_\ell(\varepsilon)} \psi^\ell_\varepsilon (r) Y_{\ell, m}(\Omega) Y_{\ell, m}^* (\Omega_k) ,
\\
\label{CoulombPhase}
& \sigma_\ell(\varepsilon) = \arg \Gamma \left( \ell + 1 +
i \frac{\hbar}{a_0 \sqrt{2m \varepsilon}} \right) ,
\end{align}
\end{subequations}
where the real radial functions $\psi^\ell_\varepsilon(r)$ are energy normalized, and $a_0$ is the Bohr radius.

The contribution to the doubly differential ionization probability from times beyond the duration of the interaction with the laser field now becomes
\begin{align}
\label{ContributionSpherical}
\left. \frac{\partial^2 P_\mathrm{ion}}{\partial \varepsilon \partial \Omega_k}\right|_\mathrm{after} & =
2 \mathrm{Im} \, \sum_{\ell, \ell'} \sum_{n, n'}
\frac{c^\ell_n [c^{\ell'}_{n'} ]^*}
{\epsilon^{\ell}_{n} - [\epsilon^{\ell'}_{n'}]^*} \\ \nonumber & \times
\langle \varphi_{\varepsilon, \Omega_k} | \gamma | \phi^\ell_n \rangle
\langle \phi^{\ell'}_{n'} | \varphi_{\varepsilon, \Omega_k} \rangle ,
\end{align}
where the sum over $n_{(\ell)}$ is still restricted to terms for which the imaginary part of $\epsilon^{\ell}_n$ is strictly negative.
The projections
\begin{align}
\label{PsiProjectionSpherical}
&
c^\ell_n = \langle \tilde{\phi}^\ell_n| \Psi(t=T) \rangle =
\int_0^\infty [\tilde{\phi}^\ell_n(r)]^* f_\ell(r; t=T) \, \mathrm{d}r \\
& = ( \tilde{\phi}^\ell_n | f_\ell(t = T) ),
\end{align}
where we, for convenience, have introduced the short-hand notation
\begin{equation}
\label{ShortHand}
( \varphi | \psi ) \equiv
\int_0^\infty [\varphi(r)]^* \psi(r) \, \mathrm{d}r .
\end{equation}
Here, $\tilde{\phi}^\ell_n(r)$ is the radial part
of the state $\tilde{\phi}^\ell_n$.
Since our CAP function $\gamma(r)$ is isotropic, these eigenstates factor into a radial part multiplied by the corresponding Spherical Harmonic.

Moreover, with a spherically symmetric local CAP, Eq.~(\ref{LocalCAP}), and Eq.~(\ref{ScatteringCoulomb}), we have
\begin{equation}
\label{ProjectScatteringNHstateSpherical}
\langle \varphi_{\varepsilon, \Omega_k} |\gamma | \phi^\ell_n \rangle =
i^{-\ell} e^{-i\sigma_\ell(\varepsilon)} Y_{\ell, m}(\Omega_k)
\left(\psi^\ell_\varepsilon | \gamma | \phi^\ell_n \right)
\end{equation}
where
\begin{equation}
\label{ShortHand2}
\left(\psi^\ell_\varepsilon | \gamma | \phi^\ell_n \right)
=
\int_{R_c}^\infty [\psi^\ell_\varepsilon(r)]^* \gamma(r) \phi^\ell_n(r)
\, \mathrm{d}r ,
\end{equation}
cf., Eq.~(\ref{ShortHand}).

All in all, Eq.~(\ref{ContributionSpherical}) may be written
\begin{align}
\label{ContributionSphericalV2}
& \left. \frac{\partial^2 P_\mathrm{ion}}{\partial \varepsilon \partial \Omega_k}\right|_\mathrm{after}  =
2 \mathrm{Im} \, \sum_{\ell, \ell'} \sum_{n, n'}
\frac{1}
{\epsilon^{\ell}_{n} - [\epsilon^{\ell'}_{n'}]^*}
\\ \nonumber & \times
(\tilde{\phi}^\ell_n | f_\ell(T)) ( f_{\ell'}(T) | \tilde{\phi}^{\ell'}_{n'})
\\ \nonumber & \times
i^{-(\ell-\ell')}e^{-i(\sigma_\ell(\varepsilon)-\sigma_{\ell'}(\varepsilon))}
Y_{\ell, m}(\Omega_k) Y^*_{\ell', m}(\Omega_k)
\\ \nonumber & \times
(\psi^\ell_\varepsilon | \gamma | \phi^\ell_n )
( \phi^{\ell'}_{n'} | \psi^{\ell'}_\varepsilon ) .
\end{align}

While this expression, admittedly, is somewhat lengthy, it simplifies considerably in case we are only interested in the ionization probability singly differential in energy. This may be determined by integrating out the ejection angle from Eq.~(\ref{DoublyDiffEnergyAngle}):
\begin{equation}
\label{IntegrateOutEjection}
\frac{\partial P_\mathrm{ion}}{\partial \varepsilon} =
\int_{4 \pi} \frac{\partial^2 P_\mathrm{ion}}{\partial \varepsilon \partial \Omega_k} \, \mathrm{d}\Omega_k .
\end{equation}
Due to the orthonormality of the Spherical Harmonics $Y_{\ell, m}^*(\Omega_k)$ in Eq.~(\ref{ScatteringCoulomb}), only terms diagonal in $\ell$ remain from the expression
in Eq.~(\ref{ContributionSphericalV2}). Moreover, the phase factors cancel so that the corresponding contribution beyond $t=T$ may be expressed as
\begin{align}
\label{SimplerExpression}
& \left. \frac{\partial P_\mathrm{ion}}{\partial \varepsilon} \right|_\mathrm{after}  =
\frac{2}{\hbar} \mathrm{Im} \, \sum_\ell
 \sum_{n, n'}
\frac{1}
{\epsilon^{\ell}_{n} - [\epsilon^{\ell}_{n'}]^*}
\\ \nonumber & \times
(\tilde{\phi}^\ell_n | f_\ell(T)) (f_\ell(T) | \tilde{\phi}^\ell_{n'}) \cdot
(\psi^\ell_\varepsilon|\gamma|\phi^\ell_n) (\phi^\ell_{n'}|\psi^\ell_\varepsilon)
\end{align}
where the sum in $n$ still only includes contributions for which $\mathrm{Im} \, \epsilon^\ell_n < 0$.

\subsection{Angle-resolved distributions}
\label{Sec_AngleResolved}

Analogously to integrating out the ejection angle from the doubly differential ionization probability, we could determine the asymptotic angular photoelectron distribution by doing the opposite:
\begin{equation}
\label{IntegrateOutEnergy}
\frac{\partial P_\mathrm{ion}}{\partial \Omega_k} =
\int_0^\infty \frac{\partial^2 P_\mathrm{ion}}{\partial \varepsilon \partial \Omega_k} \, \mathrm{d}\varepsilon .
\end{equation}
This does not, however, appear to simplify the working equations; it would seem that we are forced to calculate the angular distribution via the full, doubly differential distribution.

Thus far we have focused upon photoionization probabilities differential in asymptotic energy and ejection angle. However, it could also be interesting to take \textit{position} to be our scattering states, i.e., to impose
\begin{equation}
\label{PositionScatterinState}
| X \rangle = | r, \Omega \rangle
\end{equation}
in Eq.~(\ref{DifferentialP_generic}). This would provide a probability distribution differential in the position at which the particle is absorbed.
%
%
Specifically, the expression analogous to Eqs.~(\ref{dPdXTimeless}) that we may arrive at with the time-independent  position state, Eq.~(\ref{PositionScatterinState}), in Eq.~(\ref{DifferentialP_generic}) reads
\begin{align}
\label{DoublyDifferentialPos}
\frac{\partial^2 P}{\partial r \partial \Omega} &
=
\frac{2}{\hbar} \mathrm{Re} \, \langle r, \Omega | \eta | r, \Omega \rangle  =
\\ &
\nonumber
\frac{2}{\hbar} \mathrm{Re} \,
\int_0^t
\gamma(r) \left| \Psi(r, \Omega; t') \right|^2 \, \mathrm{d}t' .
\end{align}
Note that this time the contributions picked up at different times add \textit{incoherently}. This contrasts the situation in Eqs.~(\ref{dPdXTimeless}), in which contributions acquired at different times are added \textit{coherently}.
In other words, waves absorbed at different times by a position absorber are allowed to interfere when aggregating the momentum distribution. They are \textit{not} allowed to interfere when aggregating a \textit{position} distribution, however. This notion is crucial when it comes to the connection between a CAP and a detector. While several works
suggest such a connection in a rather generic way, it has been argued that a CAP can be seen as a detector when, and \textit{only} when, the CAP is diagonal in the basis in which it is measured~\cite{Selsto2022}. A position CAP, such as ours, acts as a detector when used to aggregate position-differential distributions, \textit{not} when it is used to obtain energy/momentum-differential probabilities.

We also note that while the position distribution in Eq.~(\ref{DoublyDifferentialPos}) is manifestly non-negative, the momentum distribution of Eqs.~(\ref{dPdXTimeless}) is not. When the absorption is too hard, it may produce ``negative probabilities'' in the momentum distribution. This artifact must, of course, be avoided in order to produce meaningful results. This may be achieved by ensuring that the CAP onset is placed sufficiently far from the interaction region. Checking for invariance in the CAP onset parameter $R_c$ is necessary to ensure numerical accuracy in any case.

Now, while the $r$-dependence of the ionization probability differential in position, Eq.~(\ref{DoublyDifferentialPos}), may not be very interesting, the angular dependence,
\begin{align}
\label{IntegrateOutR}
&
\frac{\partial P}{\partial \Omega}  =
\int_0^\infty \frac{\partial^2 P}{\partial r \partial \Omega} \, r^2 \mathrm{d}r =
\\ \nonumber &
\frac{2}{\hbar} \mathrm{Re} \, \int_0^\infty r^2 \mathrm{d}r \,
\int_0^\infty
\gamma(r) \left| \Psi(r, \Omega; t) \right|^2 \, \mathrm{d}t =
\\ \nonumber &
\frac{2}{\hbar} \mathrm{Re} \, \sum_{\ell, \ell'} Y_\ell(\Omega)
Y_{\ell'}^*(\Omega)
\int_0^t (f_{\ell'}(t')| \gamma | f_\ell(t')) \, \mathrm{d}t' ,
\end{align}
may be worth considering. Suppose the CAP represents a position-detector placed in extreme vicinity of the atom. In such a scenario, the angle $\Omega$ represents the direction from the nucleus to the point at which the particle was detected. This is not necessarily the same direction as the \textit{asymptotic} ejection angle $\Omega_k$. These directions should, however, coincide in the limit that the CAP/detector is moved towards the asymptotic region.
While this would usually be the case in an actual experiment, it would still be interesting to study to what extent the angular distributions $\partial P/d \Omega$ in Eq.~(\ref{IntegrateOutR}) and $\partial P_\mathrm{ion}/d \Omega_k$ in Eq.~(\ref{IntegrateOutEnergy}) converge towards each other as the detector is moved further away -- at least from a theoretical point of view.

If we, as in Eq.~(\ref{SumBeforeAfter}), partition the distribution of Eq.~(\ref{IntegrateOutR}) in contributions obtained during the pulse and after the pulse, the latter term would read
\begin{align}
\nonumber
\left. \frac{\partial P}{\partial \Omega} \right|_\mathrm{after} & =
\frac{2}{\hbar} \, \mathrm{Im} \,
\sum_{\ell, \ell'} Y_{\ell, m}(\Omega) Y_{\ell', m}^*(\Omega)  \, \sum_{n,n'}\frac{1}{\epsilon^\ell_n - [\epsilon^{\ell'}
_{n'}]^*}
\\ &
\label{AngularDistAfter}
\times (\tilde{\phi}^\ell_n | f_\ell(T)) (f_{\ell'}(T)| \tilde{\phi}^{\ell'}_{n'}) \cdot
(\phi^{\ell'}_{n'}| \gamma | \phi^\ell_n) .
\end{align}
We arrive at this expression by inserting $\eta_\infty$ from Eq.~(\ref{EtaIntegrated}) into Eq.~(\ref{IntegrateOutR}). As usual, we only include terms for which $\mathrm{Im} \, \epsilon^\ell_n$ is strictly negative.

\subsection{Numerical details}
\label{Sec_Numerics}

While the angular dependence of our wave function is expressed in terms of a truncated sum over Spherical Harmonics, Eq.~(\ref{ExpandSpher}), the radial dependence is approximated using a uniform grid consisting of $N$ points. In this way, by taking the partial waves $f_\ell(r; t)$ to be column vectors, the wave function is given by a matrix of dimension $N \times(L+1)$:
\begin{subequations}
\label{PsiAsMatrix}
\begin{align}
& \Psi \rightarrow F = \left( {\bf f}_0(t), {\bf f}_1(t), \cdots , {\bf f}_{L}(t) \right),
\\ &
\label{flDef}
{\bf f}_\ell(t) = \left(f_\ell(r_1; t) ,  f_\ell(r_2; t), \cdots , f_\ell(r_N; t) \right)^T ,
\\ &
\label{RadGrid}
r_n = n \cdot h, \quad h = \frac{R}{N+1} ,
\end{align}
\end{subequations}
where $R$ is the extension of the numerical domain.
In this way, the action of the Hamiltonian, Eq.~(\ref{HamHerm}), may be written as a sum of matrix multiplications -- left, right or both -- in a manner which exploits the sparsity of the problem. The differentiations involved in the kinetic energy operator and in the interaction with the external field are approximated by finite difference schemes which enforces the Dirichlet boundary conditions.

The time-evolution for $t \in [0, T]$ is implemented by means of a split operator technique in which we separate between the Hermitian and the anti-Hermitian part of the effective Hamiltonian, Eq.~(\ref{HwithCAP}):
\begin{align}
\nonumber
\Psi(t + \Delta t) & = e^{-\gamma(r) \Delta t/\hbar} e^{-i H(t+\Delta t/2) \Delta t/\hbar} e^{-\gamma(r) \Delta t/ \hbar} \\
\label{SplitOp}
& \times
\Psi(t)  + O(\Delta t^3) .
\end{align}
The anti-Hermitian part, $\exp[- \gamma(r) \Delta t/\hbar]$ is particularly simple to implement as it is diagonal in both $r$ and $\ell$. The action of the Hermitian part, on the other hand, is approximated via the Arnoldi method~\cite{Arnoldi1951}, which, in effect, reduces the dimension of the vector space in question from $N \cdot (L+1)$ so some much lower Krylov dimension $k_\mathrm{dim}$.

When it comes to the analysis of the absorbed components, we continue to take advantage of the matrix representation of $\Psi$, Eq.~(\ref{PsiAsMatrix}), also when aggregating outgoing waves. For the Coulomb waves, $\psi^\ell_\varepsilon(r)$ in Eq.~(\ref{ScatteringCoulomb}), we have made use of the MATLAB toolbox entitled \textit{Special Functions in Physics}~\cite{Schweizer2021}.

Assuming that the absorber is strong enough to attenuate all outgoing waves before they hit the boundary at $r=R$, we have, as mentioned, used the countable set of box-normalized states for the eigenstates $|\phi^\ell_n \rangle $ of the non-Hermitian Hamiltonian $H_\mathrm{eff}^{(0)}$.
If we, for each $\ell$, set up the radial part of the (reduced) wave function as columns of a matrix $P_\ell$,
the corresponding set of bi-orthogonal eigenstates, $| \tilde{\phi}^\ell_n \rangle$, are found as the columns of the matrix $\tilde{P_\ell}$ where
\begin{equation}
\label{PsiTildeAsMatrix}
\tilde{P}_\ell = \frac{1}{h} \left(P_\ell^{-1} \right)^\dagger ,
\end{equation}
cf.~Eq.~(\ref{BiOrthogonal}).

The CAP function we have used is a square monomial,
\begin{equation}
\label{CAPsquare}
\gamma(r) = \left\{ \begin{array}{ll} \gamma_0 (r-R_c)^2, & r>R_c \\ 0 & \text{otherwise} \end{array} \right. ,
\end{equation}
with CAP strength $\gamma_0 = 10^{-4}$~a.u. In our numerical examples,
we have applied various values for the CAP onset $R_c$.
In order to approximate the wave function adequately, we have employed a uniform spatial grid with increment $h = 0.2$~a.u. and partial waves extending up to $L=12$ in the $\lambda = 400$~nm case and $L=7$ in the $\lambda = 200$~nm case, cf., Eqs.~(\ref{PsiAsMatrix}). A box size of $R=150$~a.u. was more than sufficient for most of the calculations. However, for the highest value of $R_c$, $R=200$~a.u. was used.

We have used a Krylov dimension of $k_\mathrm{dim}=20$ and a numerical time step $\Delta t$ corresponding to up to 1000 steps per optical cycle.

Several expressions involve sums for which only complex ``energies'' with strictly negative imaginary parts are to be included; specifically, this applies to Eqs.~(\ref{EtaIntegrated_v2}), (\ref{DefAfter}), (\ref{ContributionSpherical}), (\ref{ContributionSphericalV2}), (\ref{SimplerExpression}),  and (\ref{AngularDistAfter}). Now, from a numerical point of view, one may rightfully question to what extent this restriction should be taken literally. As the eigenenergies of the numerical effective Hamiltonian, Eq.~(\ref{HwithCAP}), are only non-positive up to machine accuracy, the results may be sensitive to inaccuracies in the imaginary parts close to zero -- and to the particular choice of box size $R$. Moreover, while the CAP constitutes the full anti-Hermitian part of the effective Hamiltonian mathematically, this need not be so for a numerical approximation to the Hamiltonian.
In case the the Hermitian part, Eq.~(\ref{HamHerm}), is approximated in a manner which carries a certain anti-Hermitian contribution, as is the situation in our particular implementation, the distinction is blurred even further.
Thus, it may be necessary to impose a slightly more strict criterion,
\begin{equation}
\label{CutoffFinite}
\mathrm{Im} \, \epsilon_n < - c,
\end{equation}
for some positive, small $c$. This, in turn, would mean running the risk of introducing another parameter-dependency in the physical results -- analogous to the choice of the upper integration limit $t$ in Eq.~(\ref{dPdK}). On the other hand, setting $c=0$ may, in addition to the issue discussed above, be particularly problematic in the context of a long-range potential such as the Coulomb potential since also bound states, mathematically, have a finite overlap with the CAP and, thus, also a small but finite negative imaginary component.
%

Note that the cutoff introduced in Ineq.~(\ref{CutoffFinite}) will, in effect, modify the actual absorption after the interaction with the pulse as compared to the absorption which happens during the interaction -- at least in principle. Removing the restriction of Ineq.~(\ref{CutoffFinite}) all together, on the other hand, would likely render the numerical calculation unstable -- as is likely to be the case if we actually tried to propagate a numerical wave function towards extremely large final times as well.
%
We will discuss these issues when presenting the results in Sec.~\ref{Sec_Results}.

\section{Results}
\label{Sec_Results}

Our results pertain to a hydrogen atom exposed to a linearly polarized vector potential within the dipole approximation. Specifically, this laser pulse is represented by the vector potential
\begin{equation}
\label{VectorPot}
{\bf A}(t) = \left\{
\begin{array}{ll}
\hat{\bf z}\frac{E_0}{\omega} \sin^2(\pi/T \cdot t) \cos(\omega t), & 0 \leq t \leq T \\
0, & \text{otherwise}
\end{array} \right.
\end{equation}
where the pulse duration $T$ corresponds to an integer number of optical cycles; $T = N_\mathrm{cycl} \cdot 2 \pi/\omega$. The case given most emphasis in the following has the central angular frequency $\omega=0.114$~a.u., which corresponds to a photon wavelength of $400$~nm. The peak electric field strength $E_0 = 0.075$~a.u., which corresponds to a peak intensity of $2 \cdot 10^{14}$~W/cm$^2$, and the laser pulse lasts $N_\mathrm{cycl}=10$ optical cycles.

We will, in Sec.~\ref{sec_dPdOmega}, also study angular distributions for a laser pulse residing in the ultraviolet region, for which we will take $\omega$ to correspond to $200$~nm and set $E_0=0.1$~a.u. Also in this case the pulse duration is set to $10$ optical cycles.

\subsection{Photoelectron spectra}
\label{sec:AsymptoticResults}

\begin{figure}
    \centering  \includegraphics[width=1\linewidth]{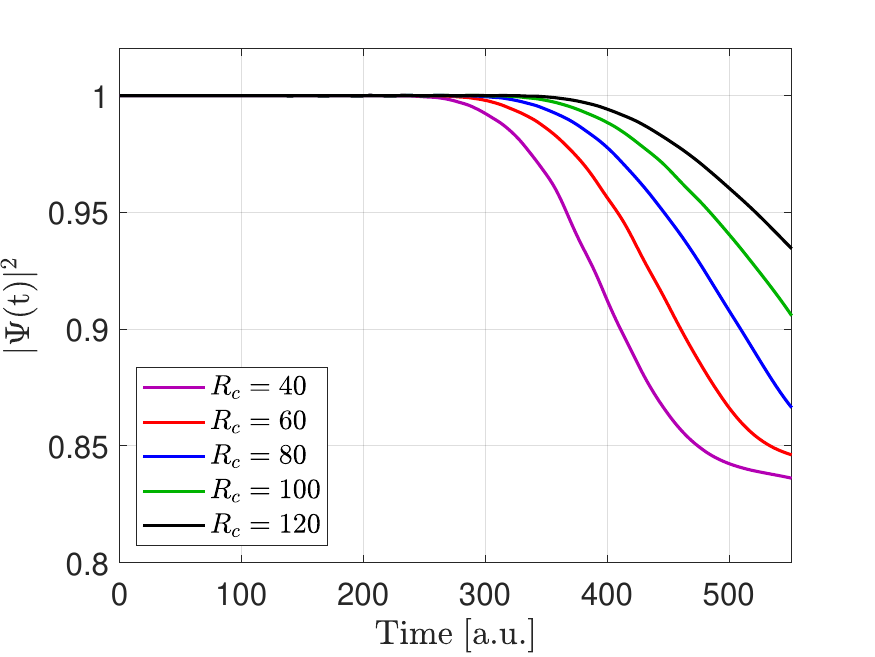}
    \caption{The norm of the wave function, $|\Psi(t)|^2$, as a function of time for various choices of the CAP onset $R_c$ as a function of time during the interaction with the laser pulse. In all cases, a significant part of the wave function as been absorbed before the pulse is over.}
    \label{fig:PabsTime}
\end{figure}

\begin{figure}
\centering
\begin{tabular}{c}
    $R_c = 40$~a.u. \\
    \includegraphics[width=.57\linewidth]{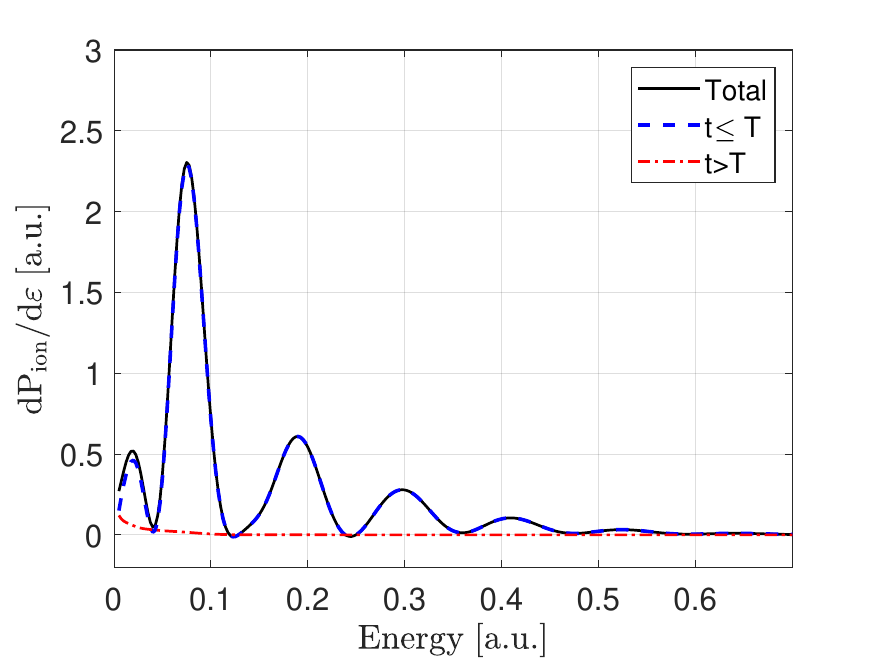} \\
    $R_c = 60$~a.u. \\
    \includegraphics[width=.57\linewidth]{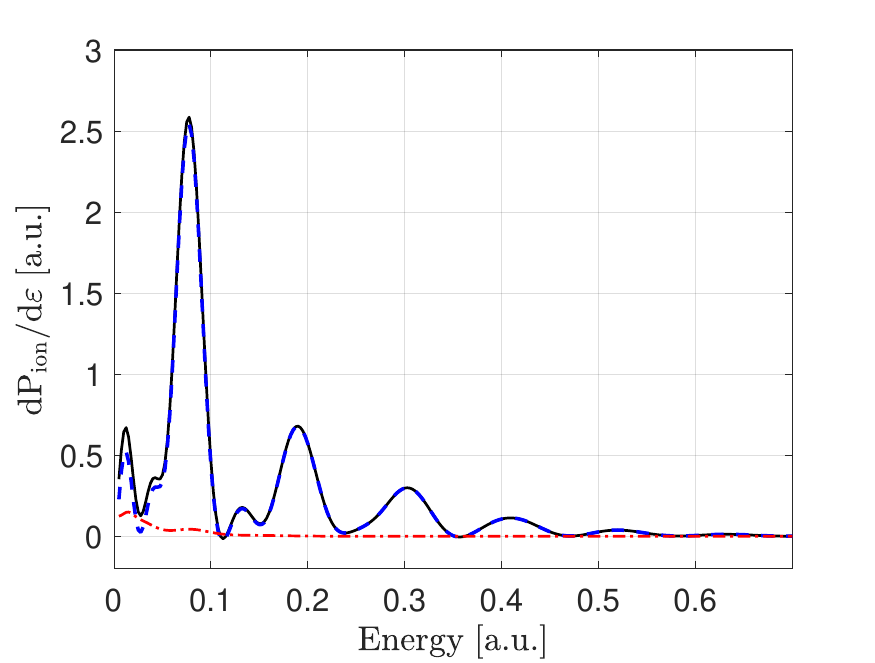} \\
    $R_c = 80$~a.u. \\
    \includegraphics[width=.57\linewidth]{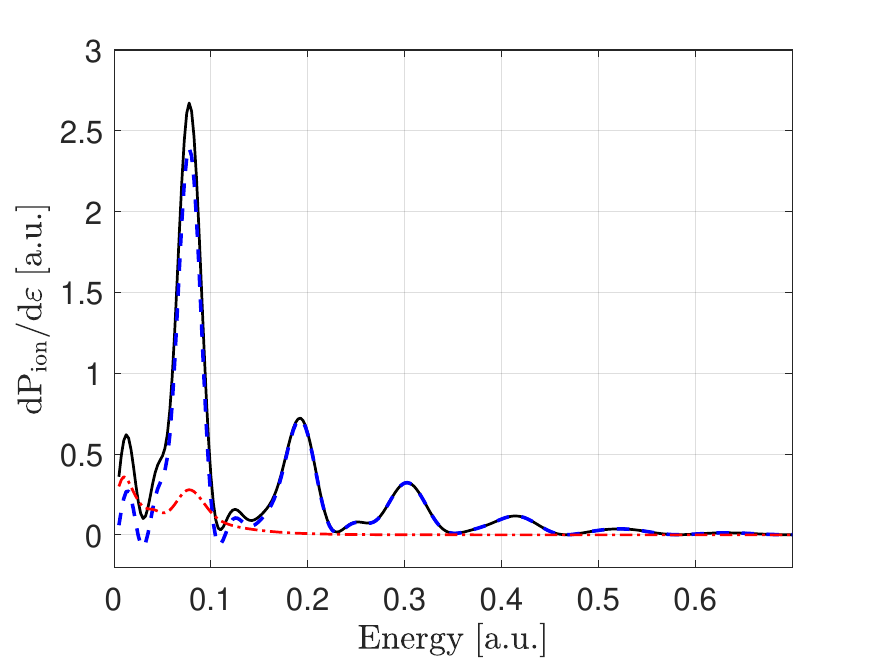} \\
    $R_c = 100$~a.u. \\
    \includegraphics[width=.57\linewidth]{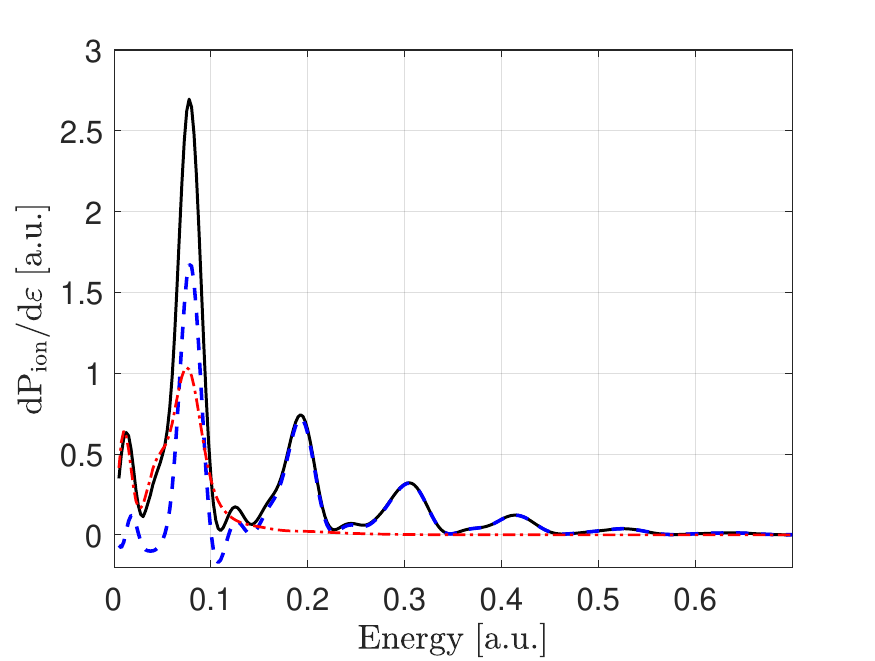} \\
    $R_c = 120$~a.u. \\
    \includegraphics[width=.57\linewidth]{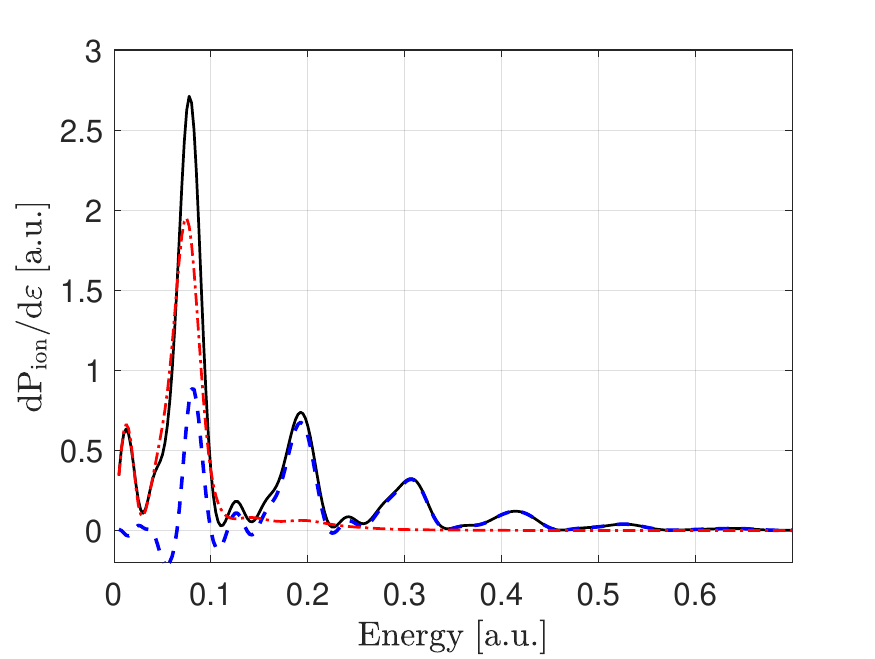}
\end{tabular}
\caption{All panels display the energy distribution of the photoelectron. They are obtained using different values for the CAP onset parameter $R_c$.
In addition to the total energy distribution (black curve), we have also displayed contributions picked up during (blue, dashed curve) and after (red, dashed-dotted curve) the interaction with the laser pulse.}
\label{fig:dPdEfigs_Rc}
\end{figure}

In Sec.~\ref{sec:PESCADO_Auto}, in Eq.~(\ref{SumBeforeAfter}) specifically, we introduced a demarcation between outgoing waves picked up during interaction with the laser field and waves absorbed afterwards. As illustrated in Fig.~\ref{fig:PabsTime}, the closer the CAP is placed to the interaction region, the more of the absorption takes place during interaction. This, in turn, means that both $\left. \partial P_\mathrm{ion}/\partial \varepsilon \right|_\mathrm{before}$ and $\left. \partial P_\mathrm{ion}/\partial \varepsilon \right|_\mathrm{after}$ will depend strongly on $R_c$ -- but, hopefully, not their sum. In Fig.~\ref{fig:dPdEfigs_Rc} we have illustrated both these partial contributions along with their sum for five different $R_c$ values distributed uniformly from $40$~a.u. to $120$~a.u.
While the contributions to the photoelectron energy distributions aggregated during interaction and after interaction differ, their total contribution does, in fact, seem to depend very weakly on $R_c$.

\begin{figure}
\centering
\begin{tabular}{c}
    \includegraphics[width=.98\linewidth]{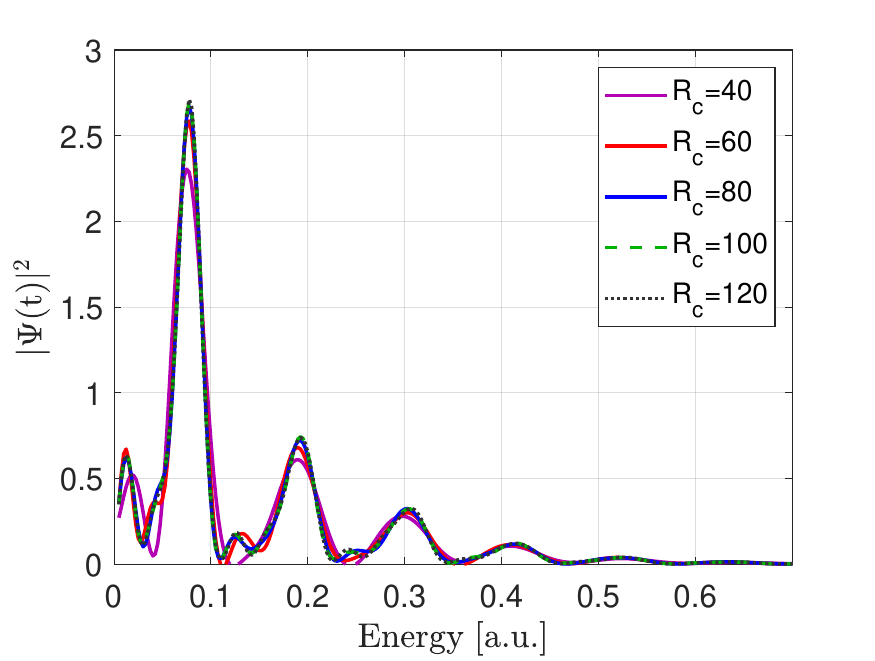} \\
    \includegraphics[width=.98\linewidth]{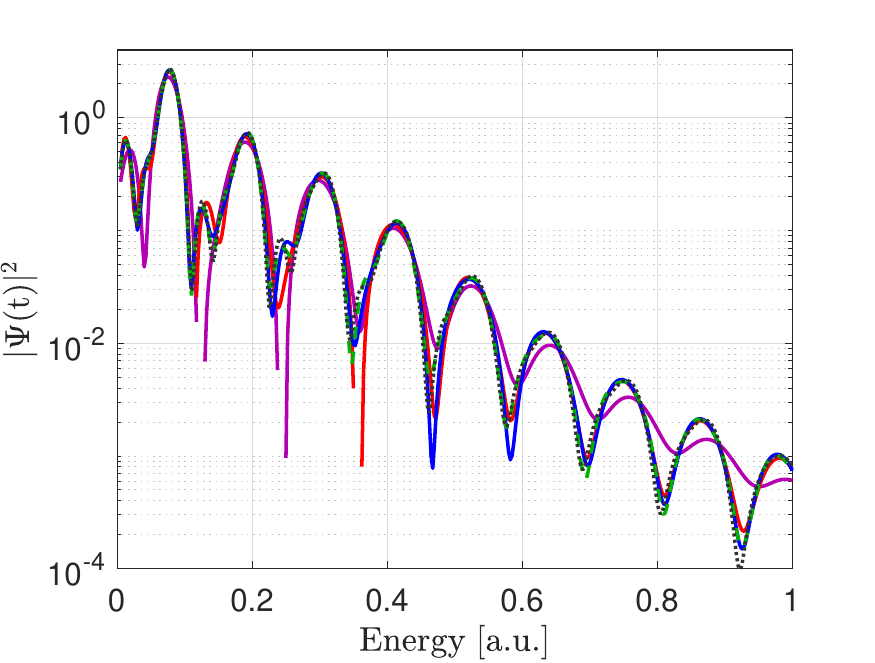}
\end{tabular}
\caption{The ionization probability differential in energy for various CAP onsets $R_c$, ranging from $40$ to $120$~a.u. The spectra are shown using linear axes in the upper panel and with a logarithmic $y$-axis in the lower one.}
\label{fig:dPdEfigs_all}
\end{figure}

This notion is confirmed in Fig.~\ref{fig:dPdEfigs_all}, which shows the total energy distributions for the $R_c$ values in Fig.~\ref{fig:dPdEfigs_Rc} plotted together. The approximations to the photoelectron spectra display an increasing degree of agreement as the CAP is moved outwards. We do, however, see a a rather strong degree of convergence already at $R_c=60$~a.u.
As demonstrated in the lower panel, in which we use a logarithmic $y$-axis, this also applies for higher \textit{above threshold ionization} peaks.

In all panels displayed in Figs.~\ref{fig:dPdEfigs_Rc} and \ref{fig:dPdEfigs_all}, we have set $c=10^{-12}$~a.u. in Ineq.~(\ref{CutoffFinite}). At lower $c$-values, we see some fluctuations at lower energies in certain cases, while it is virtually $c$-independent over several orders of magnitude beyond $10^{-12}$~a.u.

\begin{figure}
    \centering
    \begin{tabular}{lr}
    \includegraphics[width=2. cm]
    {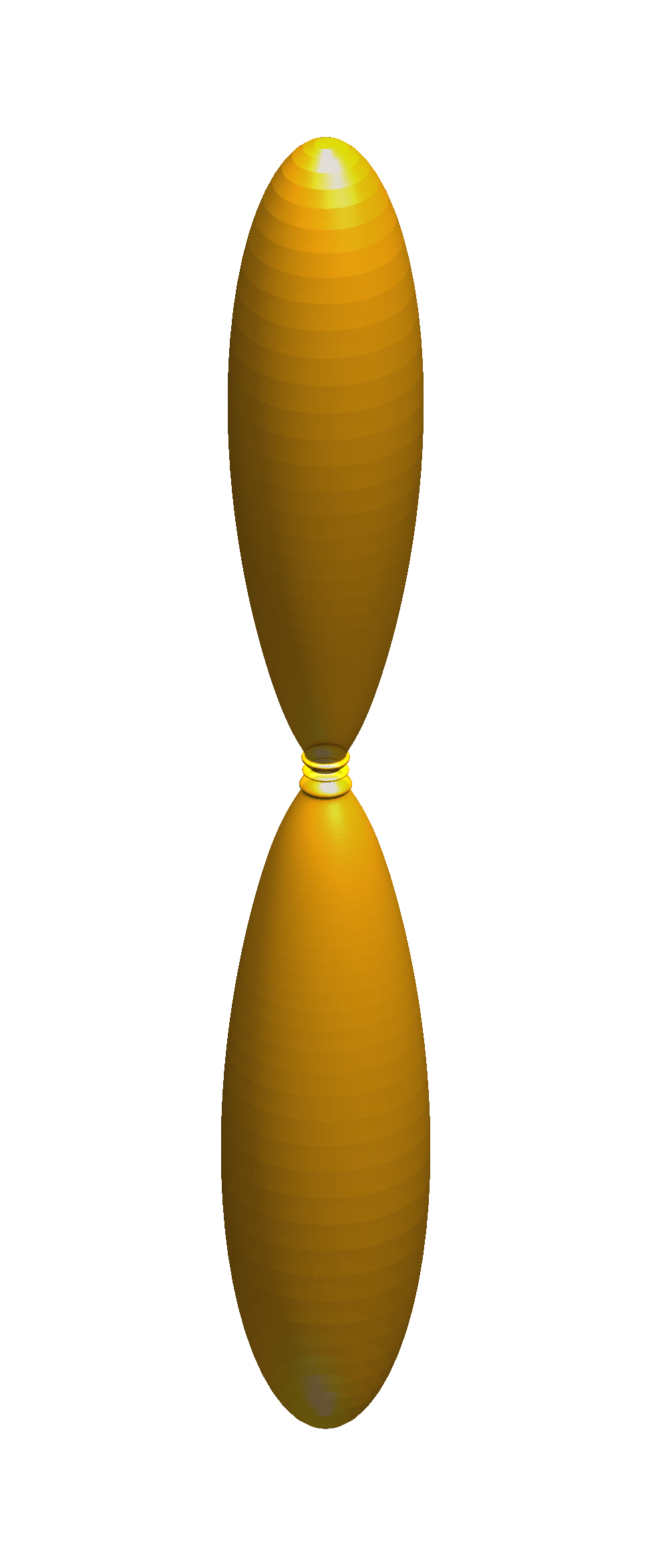} &
    \includegraphics[width=5.5 cm]
    {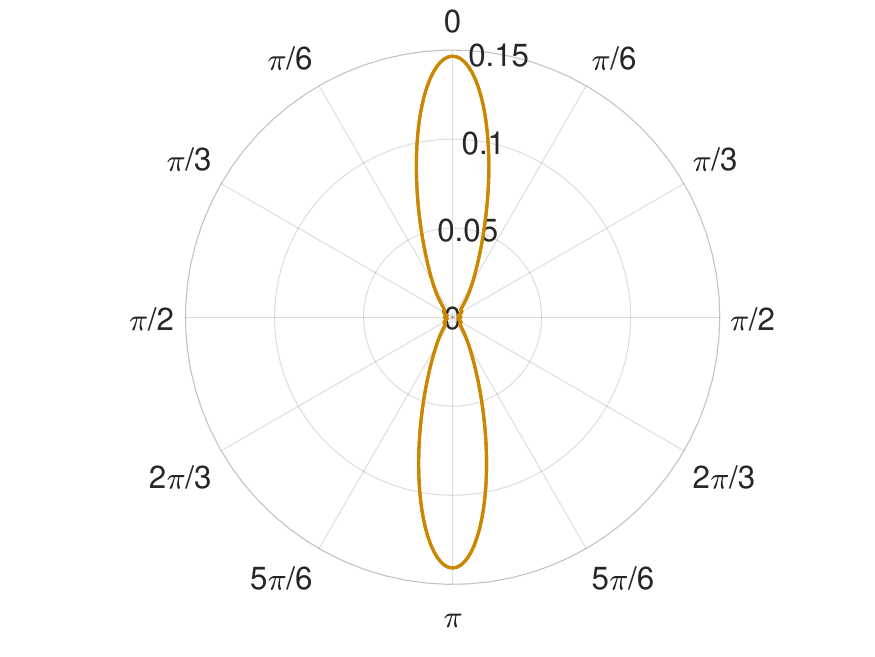}
    \end{tabular}
    \caption{The ionization probability differential in ejection angle, $\Omega_k$. To the left it is visualized as a three dimensional distribution while the right panel shows the distribution, which is independent of the azimuthal angle in the dipole approximation, in the polar angle $\theta_k$. For this particular case, the CAP onset $R_c=60$~a.u.}
    \label{fig:EjectionAngle}
\end{figure}

When it comes to the probability distribution in terms of ejection angle, the coincidence is even better. Fig.~\ref{fig:EjectionAngle} shows the ionization probability differential in the polar ejection angle $\theta_k$. It is obtained using Eq.~(\ref{IntegrateOutEnergy});  due to the cylindrical symmetry of the system, the distribution is independent of the azimuthal ejection angle $\varphi_k$. Still, we have demonstrated the angular distribution both as a three-dimension visualization and as a function of the polar angle $\theta_k$ alone. For this particular case, the CAP onset is $60$~a.u. We have not displayed the corresponding results for any of the other $R_c$ values used here simply because they would be virtually indistinguishable.

\begin{figure}
\centering
$40$~a.u. \hspace{.75cm} $60$~a.u. \hspace{.75cm} $80$~a.u. \hspace{.3cm} $100$~a.u.
\includegraphics[width=1\linewidth]{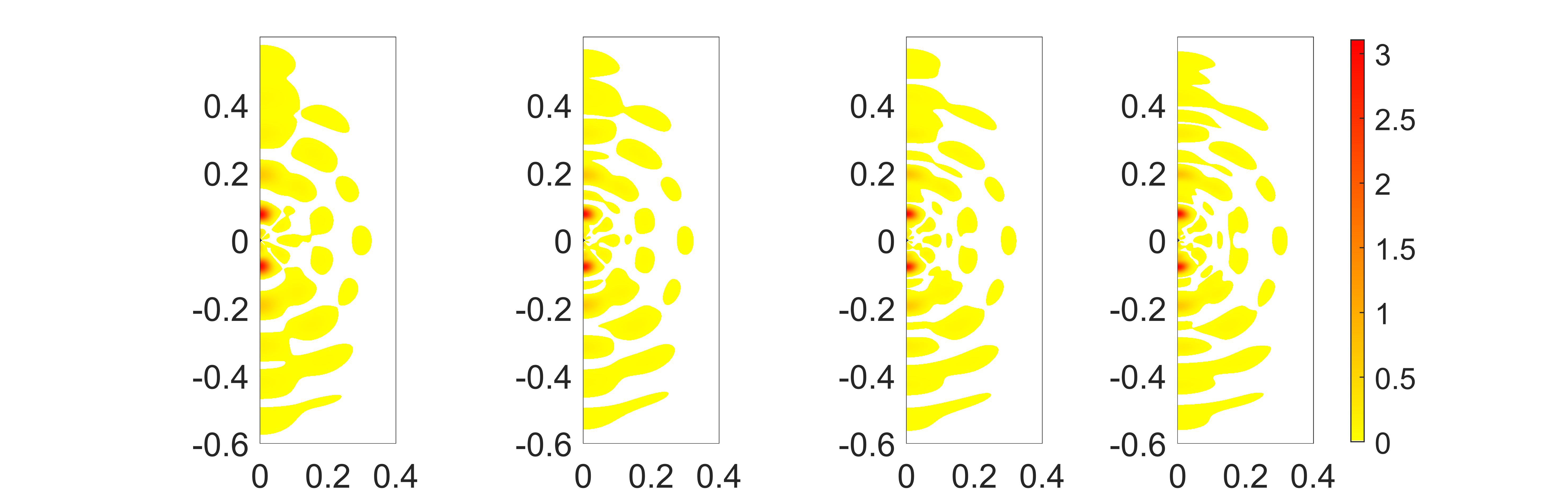}
\caption{The doubly differential ionization probability obtained using various choices for the CAP onset $R_c$. Specifically, from left to right, $R_c= 40, 60, 80$ and $100$~a.u. The $x$ and $y$-axis correspond to $\varepsilon \cos \theta_k$ and $\varepsilon \sin \theta_k$, respectively, in atomic units.}
\label{fig:DoublyDiff}
\end{figure}

With coincidence in both energy and ejection angle, cf. Figs.~\ref{fig:dPdEfigs_all} and \ref{fig:EjectionAngle}, it should come as no surprise that also the predicted doubly differential ionization probability features a very weak $R_c$-dependence. This is illustrated in Fig.~\ref{fig:DoublyDiff}. Although, as we may see from Fig.~\ref{fig:EjectionAngle}, the ejected electron to a large extent follows the polarization of the laser pulse, we may still make out circles corresponding to above threshold ionization.

\subsection{Distributions in absorption angle}
\label{sec_dPdOmega}

\begin{figure}
    \centering
    \includegraphics[width=.9 \linewidth]{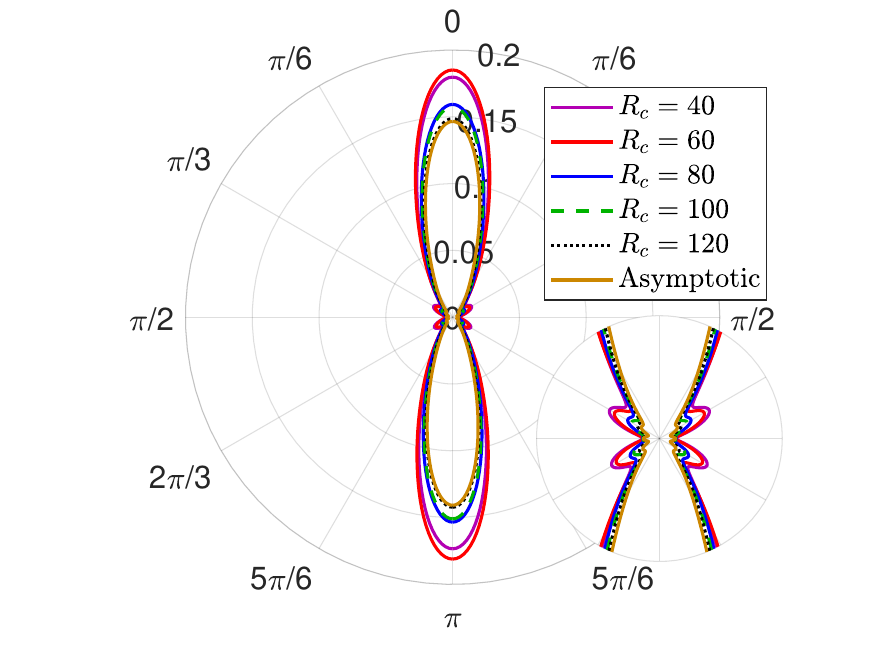}
    \caption{The ionization probability differential in absorption angle $\theta$. The distributions are shows for five different values of the CAP onset parameter $R_c$, chosen uniformly from $40$ to $120$~a.u. Also shown is the ionization probability differential in ejection angle, $\theta_k$, the same one as in Fig.~\ref{fig:EjectionAngle}. The insert is a close-up.}
    \label{fig:AbsorptionAngle1}
\end{figure}

If we, instead of the asymptotic ejection direction $\Omega_k$, consider the absorption probability differential in absorption angle, Eq.~(\ref{IntegrateOutR}), we do see a strong $R_c$-dependence. Results are shown in Fig.~\ref{fig:AbsorptionAngle1}. The color encoding corresponds to the one used in Figs.~\ref{fig:PabsTime}, \ref{fig:dPdEfigs_all} and \ref{fig:EjectionAngle}.
The fact that the situation differs so strongly from the distribution in the asymptotic ejection angle $\theta_k$
can be attributed to a number of
reasons:
\begin{itemize}
\item The absorbed waves are aggregated incoherently; any interference effects in the position distribution are ruled out after absorption.
\item This happens while the electron is still under the influence of both the Coulomb potential and the time-dependent external laser field.
\item Any part of the wave which has an appreciable overlap with the CAP would contribute; there is no inherent distinction between population of excited states and low-energy parts pertaining to the continuum.
\end{itemize}
The attentive reader may have noticed that the probability distributions of Eq.~(\ref{DoublyDifferentialPos}), and those derived from it, are not referred to as any differential \textit{ionization} probabilities. This is due to the last point above. We will return to this issue.

More importantly, all three points are consistent with the notion of the local CAP acting as a detector, as discussed in Sec.~\ref{Sec_AngleResolved} and explained in detail in Ref.~\cite{Selsto2022}.
This incoherent aggregation renders the resulting distribution sensitive to both the time and position of absorption -- as would be the case with an actual position detector placed in the extreme vicinity of the atomic nucleus as well.

Now, one may rightly argue that in order for us to consider any distribution emerging from Eqs.~(\ref{DoublyDifferentialPos}) or (\ref{IntegrateOutR}) to correspond to the result of measurement, the position and momentum coordinates should correspond to the lab frame. Or, in other words, calculations should be performed in the length gauge -- as opposed to the velocity gauge as in the present case. However, since the position variable coincides in these two gauges, we may safely apply the velocity gauge wave formulation of the wave function;  the difference would only correspond to a time and position dependent phase factor imposed on the wave function which would cancel out in Eq.~(\ref{DoublyDifferentialPos}).

\begin{figure}
    \centering
    \includegraphics[width=.9\linewidth]{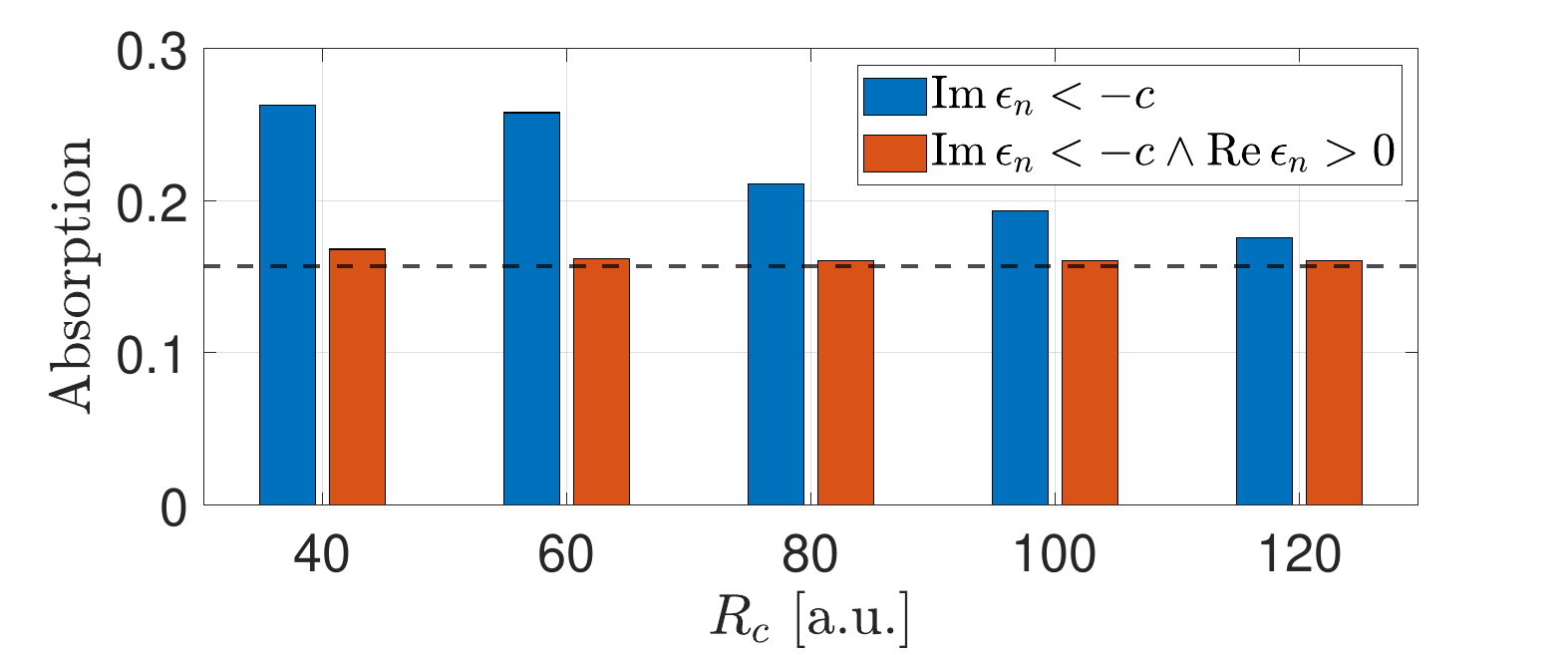}
    \caption{The total absorption obtained during the pulse duration $T$ for various values of the CAP onset $R_c$. The dashed, horizontal line is the total ionization probability estimated as the integral of the energy-distribution.}
    \label{fig:TotalAbsRc}
\end{figure}

As the absorption angle $\Omega$ and the asymptotic ejection angle $\Omega_k$ simply are different quantities -- with different probability distributions, there is no reason to expect their distributions to coincide. However,
the $\Omega$-distribution should approach $dP/d \Omega_k$ in the limit that the CAP onset $R_c$ becomes large.
The results demonstrated in Fig.~\ref{fig:AbsorptionAngle1} are consistent with this notion. In particular, the smaller lobes seen near
$\theta = \pi/2 \pm \pi/6$ in Fig.~\ref{fig:AbsorptionAngle1} are significantly smaller for the highest $R_c$ values.
The lobes seen at low values of $R_c$ seem to originate from from excited states, for which $\ell=3$ is the dominating channel. With a CAP placed further away from the nucleus, the overlap between excited states and the CAP function becomes smaller, effectively reducing their contribution towards the total absorption probability. The left columns displayed in Fig.~\ref{fig:TotalAbsRc} demonstrate this.
The histogram shows the total $\theta$-integrated absorption -- along with the ionization probability estimated by the energy-integrated distributions shown in Figs.~\ref{fig:dPdEfigs_Rc} and \ref{fig:dPdEfigs_all}. This estimate, which underestimates the ionization probability slightly due to the finite limit imposed at both ends of the energy scale, is virtually $R_c$-independent.
As $R_c$ is increased, the total absorption is reduced towards the ionization probability.

The angular distributions displayed in Fig.~\ref{fig:AbsorptionAngle1} are all obtained using the same $c$-value in Ineq.~(\ref{CutoffFinite}) as in the previous section, namely $c=10^{-12}$~a.u. Imposing such a cutoff reduces the contribution from excited states, thus also reducing the total absorption for $t>T$. Contrary to probability distributions differential in energy and/or asymptotic ejection angle, which are obtained by projection onto the proper scattering states, the angular distributions shown in Fig.~\ref{fig:AbsorptionAngle1} are quite sensitive to $c$. Specifically, converged results were hard to obtain using $c=0$. In general, while reasonably converged results are obtained using the reported numerical parameters also in this case using $c=10^{-12}$~a.u., the degree of convergence is weaker than for the numerical results presented in Sec.~\ref{sec:AsymptoticResults}. Numerical issues are manifested in the fact that the left and right eigen values of the effective Hamiltonian may differ numerically, cf. Eq.~(\ref{LeftEigenStates}). Moreover, the numerical matrix of eigenvectors of the non-Hermitian effective Hamiltonian, $P_\ell$, was not always well conditioned. Consequently, we have replaced full inversion in Eq.~(\ref{PsiTildeAsMatrix}) by the Moore-Penrose pseudoinverse when calculating absorption probabilities differential in absorption angle.

When the calculations involving projection onto scattering states do not suffer from these numerical issues to the same degree, this may be related to the fact that the projections allow us to exclude low energy components. The fact that the energy grid we project onto has a finite lower limit, allows us to evade issues induced by the overlap between the CAP and both Rydberg states and outgoing waves with near-zero energy. As discussed, when applying Eq.~(\ref{AngularDistAfter}), such filtering is less straight forward.

\begin{figure}
    \centering
    \includegraphics[width=.9\linewidth]{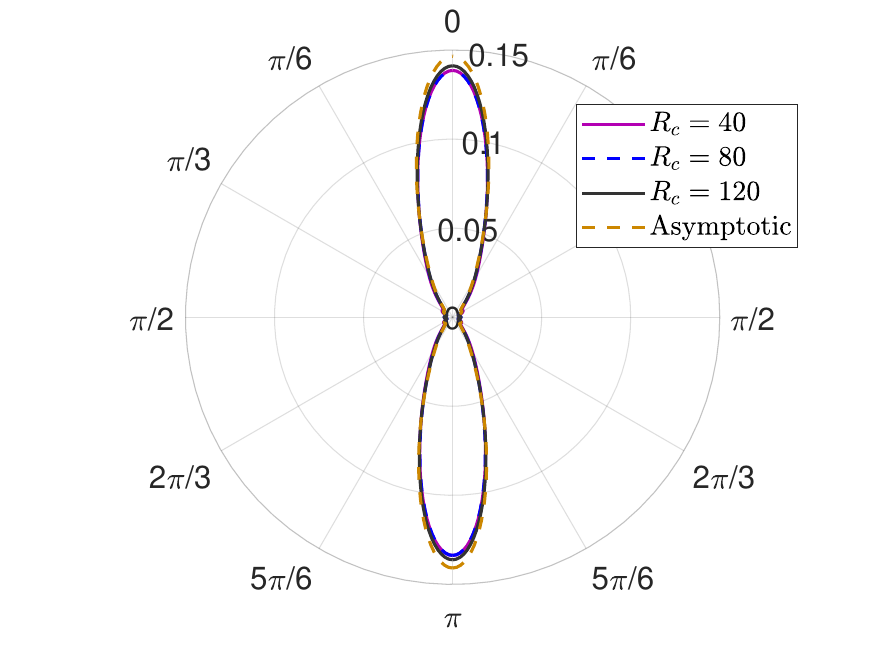}
    \caption{Probability distributions differential in absorption angle subject to the condition that $\mathrm{Re} \, \epsilon^\ell_n>0$ -- in addition to $\mathrm{Im} \, \epsilon^\ell_n < 0$ in Eq.~(\ref{AngularDistAfter}). Results are calculated with CAP onset parameter $R_c=40$, $80$ and $120$~a.u. -- and compared with the ionization probability differential in asymptotic ejection angle.}
    \label{fig:AbsorptionAngleRealCutoff}
\end{figure}

In a heuristic attempt to reduce the significance of bound states in the angular distribution, we have restricted the sum in Eq.~(\ref{AngularDistAfter}) to only include terms for which $\mathrm{Re} \, \epsilon^\ell_n > 0$ -- in addition to imposing Ineq.~(\ref{CutoffFinite}). The resulting total absorption is displayed in the right columns in the histogram in Fig.~\ref{fig:TotalAbsRc}.
Indeed, these ``absorption probabilities'' consistently lies closer to the actual ionization probability. And the corresponding angular distributions, shown in Fig.~\ref{fig:AbsorptionAngleRealCutoff}, do also resemble the ionization probability differential in ejection angle, Fig.~\ref{fig:EjectionAngle}, to a much higher degree than those in Fig.~\ref{fig:AbsorptionAngle1}. Note that the two-dimensional polar plot presented in Fig.~\ref{fig:AbsorptionAngleRealCutoff}, which does not visually account for the Jacobian factor $\sin \theta_{(k)}$, may be somewhat deceiving in the sense that the total yield may appear to be larger for the distribution in asymptotic ejection angle. In fact, as can be seen in Fig.~\ref{fig:TotalAbsRc}, also when we restrict the sum in Eq.~(\ref{AngularDistAfter}) to include only complex energies for which the real part is positive, the total absorption still exceeds the total yield obtained when projecting onto scattering states.

\begin{figure}
    \centering
    \begin{tabular}{c}
    \includegraphics[width=.9 \linewidth]{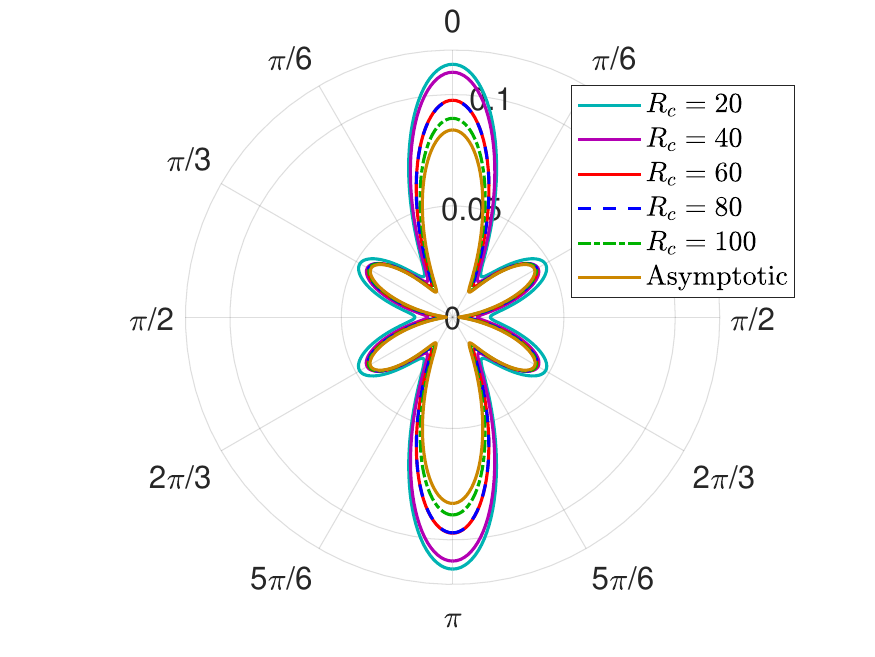} \\
    \includegraphics[width=.9 \linewidth]{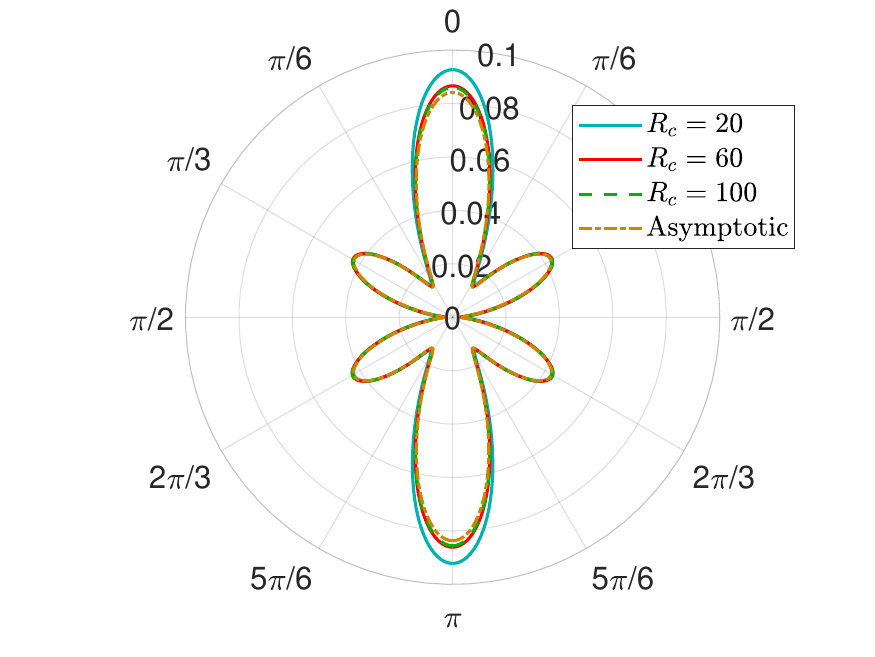}
    \end{tabular}
    \caption{Angular probability distributions pertaining to a hydrogen atom exposed to a 10-cycle laser pulse with $\omega = 0.228$~a.u. and $E_0=0.1$~a.u. The upper panel displays the probability differential in absorption angle, Eq.~(\ref{IntegrateOutR}), with the cutoff parameter $c$ set to $10^{-12}$~a.u., cf. Ineq.~(\ref{CutoffFinite}) and Eq.~(\ref{AngularDistAfter}). In the lower panel the additional restriction that $\mathrm{Re} \, \epsilon^\ell_n > 0$ in Eq.~(\ref{AngularDistAfter}) has been imposed.}
    \label{fig:200nmAngular}
\end{figure}

As mentioned, we have also studied angular distributions for a case corresponding to a photon wavelength of $200$~nm. A very similar case was studied in~\cite{Dalen2025} -- with emphasis on the photoelectron \textit{energy} spectra. In Fig~\ref{fig:200nmAngular} we have displayed results analogous to those presented in Figs.~\ref{fig:AbsorptionAngle1} and \ref{fig:AbsorptionAngleRealCutoff}. While $\theta_{(k)}=0, \pi$ are the dominant directions also in this case, lobes near $\theta_{(k)} \approx \pi/2 \pm \pi/6$ are much more prominent here. Apart from this fact, the main conclusions are the same: While the ionization probability differential in asymptotic ejection probability is virtually independent of the CAP onset parameter $R_c$, the distributions in absorption angle show a strong $R_c$-dependence. The latter distribution seems to become increasingly similar to the former one as $R_c$ increases. Finally, with the heuristically motivated restriction that the sum over $n$ in Eq.~(\ref{AngularDistAfter}) only includes complex ``energies'' with positive real part, the coincidence between the two distributions improves considerably. In this pragmatic approach, the notion of a CAP acting as a detector does not apply.

Finally we emphasize that the PESCADO method readily generalizes to systems with several particles; using second quantization, the CAP still acts as a one-particle interaction in a many-particle context~\cite{Selsto2010, Kvaal2011, Selsto2021}. For continued simulation and analysis of the reminder of the system after first absorption, it is necessary to solve a master equation in addition to the original Schr{\"o}dinger equation. Full analysis of the first ionization process does, however, come with little computational overhead. Correspondingly, for an implementation able to simulate the wave function of a many-particle system undergoing ionization, adapting the PESCADO method to study single electron ionization of many-particle systems may very well turn out to be a particularly low-hanging fruit.


\section{Conclusions}
\label{sec:Conclusions}

We demonstrated how the PESCADO method may be adapted in a manner which evades the need to propagate a wave packet long beyond the interaction with a laser pulse.
Our results demonstrate that well converged singly and doubly differential photoelectron distributions may be obtained even at hard truncation of the numerical domain.
We explained this by pointing to two features of the method. Firstly, the method is, to a large extent, able to accommodate for the long range nature of the Coulomb potential. Secondly, it allows for waves absorbed at different times to interfere in momentum space; the absorbed components are aggregated coherently.

In position space, however, this aggregation happens incoherently for a local CAP. Correspondingly, probability distributions differential in absorption angle demonstrated strong dependence on the onset, as would also be the case with a detector placed in the extreme vicinity of the atom.

\section*{Acknowledgements}

It is a pleasure to thank Prof.~Wolfgang Schweizer for providing agile and clarifying support in regard to his \textit{Special Functions in Physics} toolbox. We are also pleased and grateful for the help provided by Hedda Marie Westlin in facilitating the use of our computer resources.


\end{document}